\documentclass[reprint, amsmath,amssymb, aps,pre,floatfix,]{revtex4-2}

\usepackage{graphicx}
\usepackage{dcolumn}
\usepackage{bm}
\usepackage[position=t,singlelinecheck=off,justification=raggedright]{subfig}
\usepackage[]{natbib}
\usepackage[colorlinks,citecolor=blue]{hyperref}

\begin{document}

\title{Unconventional Rashba spin splitting and persistent spin helices based on SU(2) symmetry in PtSe$_2$ nanoribbons}

\author{Bo-Wen Yu}
\affiliation{Beijing National Laboratory for Condensed Matter Physics, Institute of Physics, Chinese Academy of Sciences, Beijing 100190, China}
\affiliation{School of Physical Sciences, University of Chinese Academy of Sciences, Beijing 100049, China}
\author{Bang-Gui Liu}\email{bgliu@iphy.ac.cn}
\affiliation{Beijing National Laboratory for Condensed Matter Physics, Institute of Physics, Chinese Academy of Sciences, Beijing 100190, China}
\affiliation{School of Physical Sciences, University of Chinese Academy of Sciences, Beijing 100049, China}

\begin{abstract}
2D materials can host interesting physics and have important applications in various fields. Recent experiment shows that monolayer PtSe$_2$ nanoflakes with neutral zigzag edges are stable. Here, we study semiconducting stoichiometric PtSe$_2$ nanoribbons with the stable neutral zigzag edges (with $N$ describing different nanoribbon width) through combining detailed first-principles investigation with low-energy model analysis. Our careful analysis of first-principles conduction and valence bands (with the spin-orbits coupling taken into account) indicates that the low-energy bands assume relativistic energy dispersion in an energy window of [-0.2 eV, 0.2eV] (at least) and have large unconventional Rashba spin splitting (for even $N$). Furthermore, it is demonstrated that  the low-energy bands can be well described by an effective one-dimensional electron model and the semiconductor gap will remain finite even for large $N$. Most importantly, it is shown that there exists SU(2) spin symmetry in both of the conduction and valence bands for each edge, which implies persistent spin helices (in the spin xy plane) and spin-conserving carrier transport. When the inter-edge interaction becomes weak ($N$ is large enough), a nearly-perfect Dirac fermion system can be achieved through combining the valence and conduction bands. Thus we realize  unconventional Rashba splitting, double SU(2) spin symmetry, persistent spin helices/textures, and pure Dirac fermion systems in stable monolayer PtSe$_2$ nanoribbons.
\end{abstract}

\maketitle

 \section{INTRODUCTION}

Advent of graphene has stimulated huge interest in two-dimensional (2D) materials for basic scientific exploration and practical applications\cite{rx00,rx01,rx02,st2d2020rmp,2Dtransistor1,2Dtransistor}. Transition metal dichalcogenides (TMDs) as 2D semiconductors have various amazing properties in the field of high-performance electronic, spintronic, and optoelectronic devices\cite{rx03,rx04,rx05,2Ddevice1,2Ddevice2,reviewTMD2023}. Platinum diselenide (PtSe$_2$) becomes attractive because PtSe$_2$ monolayer and multilayers can be made stable on various surfaces\cite{rx10,rx06,rx07,rx13,rx09,growth2024a,growth2024b}. Its electronic structures are investigated with first-principles methods\cite{diracsm1,ptse2-1,ptse2-2,ptse2-3,gw}. Further exploration indicates that the PtSe$_2$ materials can host many intriguing properties such as type-II Dirac semimetals\cite{weylsm1,diracsm1,diracsm2}, spin-layer locking\cite{rx11}, thickness-driven semiconductor-metal transition \cite{semi-metal}, highly-sensative piezoresistivity\cite{rx12},  air-stability\cite{rx14}, magnetism\cite{ptse2mag},  and Majorana corner modes\cite{top2024a}, and they have potential applications in electronic and optoelectronic devices\cite{rx15,rx16,rx17,rx18,rx19,ptse2-device,growth2024a,growth2024b}. Very importantly, recent experiment shows that  monolayer PtSe$_2$ nanoflakes with a neutral zigzag edge (terminated with one Se atom per unit cell) are stable\cite{rx09}. Stoichiometric monolayer PtSe$_2$ nanoribbons can be constructed with this neutral single-Se termination only (for both edges), in comparison to usual stoichiometric nanoribbons with Pt and double-Se terminations\cite{reviewTMD2023}. First-principles electronic structures are calculated for PtSe$_2$ nanoribbons with the single-Se termination \cite{rx09,abc2,abc3,abc4,abc5,top2024a} and other terminations\cite{abc1,abc2,abc4,abc5}. 
Although these have been achieved, it is still highly desirable to explore more effects and insight of monolayer PtSe$_2$ nanoribbons with stable zigzag edges\cite{rx09}.

Rashba effect due to the spin-orbits coupling can substantially change the parabolic dispersion of band edges of semiconductors, causing various phenomena and applications\cite{rashba,bitei,rashba2,rashba3,rashbalike}.  When one of the two Rashba parameters is effectively made zero, there appears a special SU(2) spin symmetry in the Rahsba-split energy bands and a persistent spin helix (PSH)\cite{sfet2003,psh2006prl}. This PSH and enhanced spin lifetime are observed in high-mobility parabolic carrier systems in GaAs quantum wells and some 2D systems\cite{pst2017rmp,psh1qw,psh2qw,st2d2020rmp}. It is very interesting to explore such effects and surprising properties in the monolayer PtSe$_2$ nanoribbons with stable zigzag edges. On the other hand, since graphene was shown to host a 2D Dirac fermion system\cite{rx01,rx02,2Ddirac,DiracSi}, it is always very interesting to realize more Dirac fermion systems in condensed matter physics.

Here, we study the monolayer PtSe$_2$ nanoribbons with the stable neutral zigzag edges through first-principles investigation and further analyses of effective low-energy electron models. It is found that the edge-related low-energy electronic bands, in the semiconductor gap of the monolayer PtSe$_2$, can be well fit with a relativistic energy dispersion relation in the energy window of [-0.2 eV, 0.2 eV]. These low-energy conduction and valence bands can be well described by an effective electron model and its semiconductor gap remains finite. We also find large unconventional Rashba spin splitting and exact SU(2) spin symmetry in both of the conduction and valence bands for even $N$ (with $N$ describing the width of the nanoribbon). This SU(2) spin symmetry implies PSH and spin-conserving transport along each edge.  Furthermore, a nearly-perfect Dirac fermion system can be realized when the inter-edge interaction becomes weak enough (with $N$ being large enough). More detailed result will be presented in the following.

\section{METHODOLOGY}

The first-principles calculations are performed with the projector-augmented wave (PAW) method within the density functional theory\cite{rx24}, implemented in the Vienna Ab-initio simulation package software (VASP) \cite{rx25}. The generalized gradient approximation (GGA)  by Perdew, Burke, and Ernzerhof (PBE) \cite{rx26} is used as the exchange-correlation functional. The  self-consistent calculations are carried out with a $\Gamma$-centered ($12\times 1\times 1$) Monkhorst-Pack grid for the computational slab model  of the nanoribbons\cite{rx27}.  The inter-layer vacuum thickness is set to 25 \AA{} and the inter-edge vacuum separation within the plane is equal to 20 \AA. The kinetic energy cutoff of the plane waves is set to 450 eV. The convergence criteria of the total energy and atomic force are set to 10$^{-6}$ eV and 0.01 eV/\AA{}. The spin-orbit coupling (SOC) is taken into account in the band calculation and lattice structure optimization.

\begin{figure}
\includegraphics[width=1.0\columnwidth]{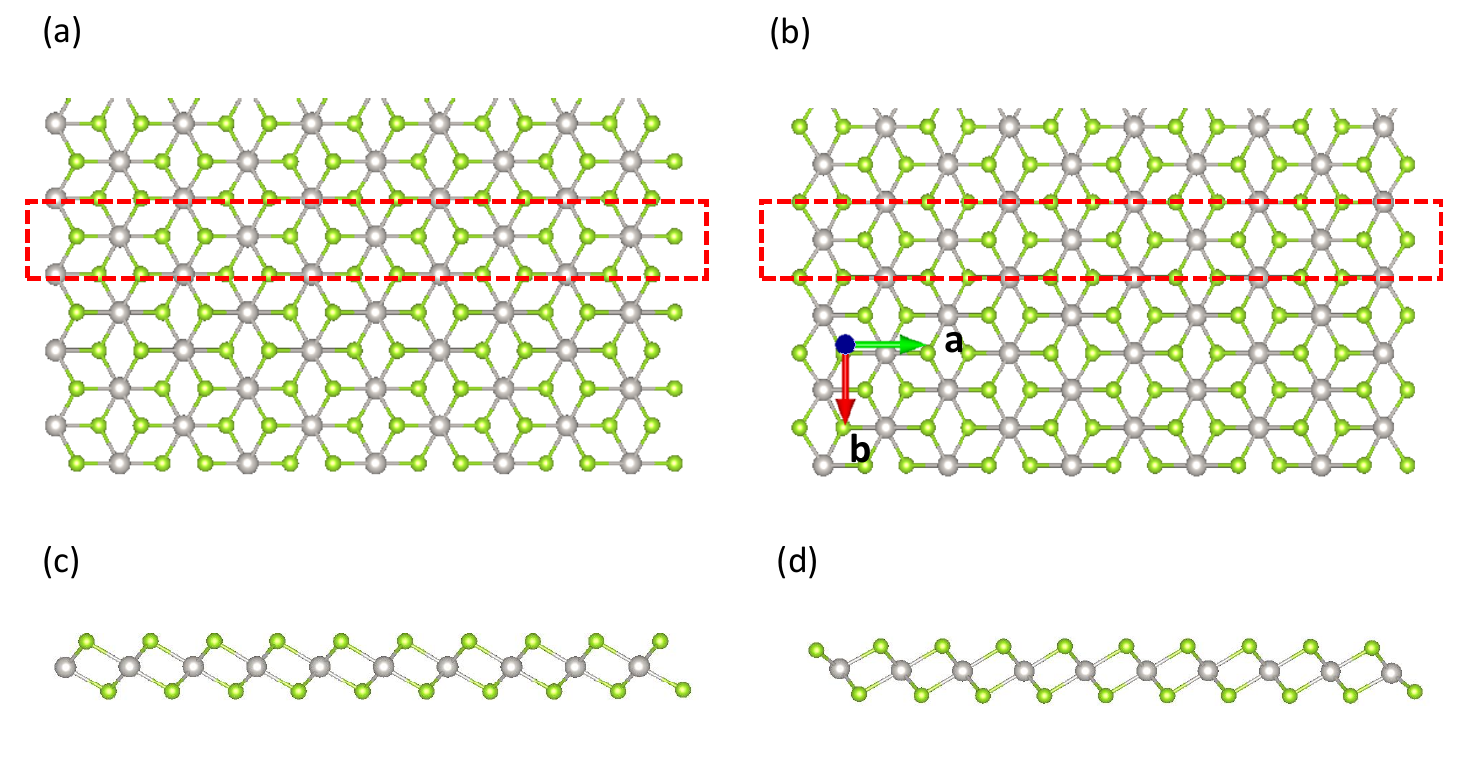}
\caption{\label{fig1} Top view (a) and side view (c) of the usual zigzag PtSe$_2$ nanoribbon with Pt and double-Se edges. Top view (b) and side view (d) of the neutral zigzag PtSe$_2$ nanoribbon with the single-Se edges. The nanoribbon is along the a axis. The top view shows the a-b plane (the upper part), and the side view the b-c plane (the lower part). }
\end{figure}

\section{RESULT AND DISCUSSION}

\subsection{Monolayer PtSe$_2$ edges}

Usually, a monolayer TMD can assume one of the two stable  monolayer structures, H-phase and T-phase. For PtSe$_2$, the the layered T-phase is the most stable, and the T-phase monolayer is a semiconductor. The T monolayer of PtSe$_2$ has C$_3$-centered symmetry and there are six Se atoms near each Pt atom in the PtSe$_2$ lattice. Similar to graphene monolayer which also has the 120$^{\circ}$ rotational symmetry, the T monolayer of PtSe$_2$ can have armchair and zizag edges. 
The armchair edge consists of Pt-Se-Se-Pt and keeps the stoichiometric ratio of PtSe$_2$ unchanged, and the zigzag edge is more complex and one can construct three types of zigzag edges: Pt, double-Se, and Single-Se edges, as shown in Fig. \ref{fig1}. The Pt and double-Se zigzag edges are similar to those of the H phase and have one Pt atom or two Se atoms per unit cell along the edge. There will be net charge on each of these two edges and they tend  instable against reconstruction. As shown in Fig. \ref{fig1} (a,c), however, the usual zigzag nanoribbons can be constructed by taking the Pt and double-Se edges as the two edges \cite{abc-exp1,reviewTMD2023}, and some magnetism can appear in such nanorribons with Pt-termination \cite{abc1,abc2,abc4,ptse2mag}. Fortunately, one can construct a zigzag nanorribon with this type of single-Se zigzag edges only, as shown  in Fig. \ref{fig1} (b,d). Electronic structures of such nonmagnetic nanoribbons are studied \cite{rx09,abc2,abc3,abc4,abc5,top2024a}. This zigzag edge keeps the correct stoichiometric ratio of PtSe$_2$ and is charge neutral, and its stability is shown experimentally\cite{rx09}.

In the following, we study the monolayer PtSe$_2$ nanoribbon with the stable neutral zigzag edges. Our computational model is shown in Fig. \ref{fig1} (b,d). We describe the width of the PtSe$_2$ nanoribbon by $N$ which is the number of the PtSe$_2$ units along the perpendicular b direction. It should be noticed that the symmetry is different between odd $N$ and even $N$. We have optimized the crystal structures with $N=3\sim10$. The atomic positions in the interior region are not changed by the optimization, and the atoms near the edges are relaxed with the optimization. The outer-most Se-Pt bond length is shortened by 2.7\% (for $N=10$), and the bond length deviation decreases rapidly when the distance from the edge increases.

\begin{figure}
\includegraphics[width=0.46\textwidth]{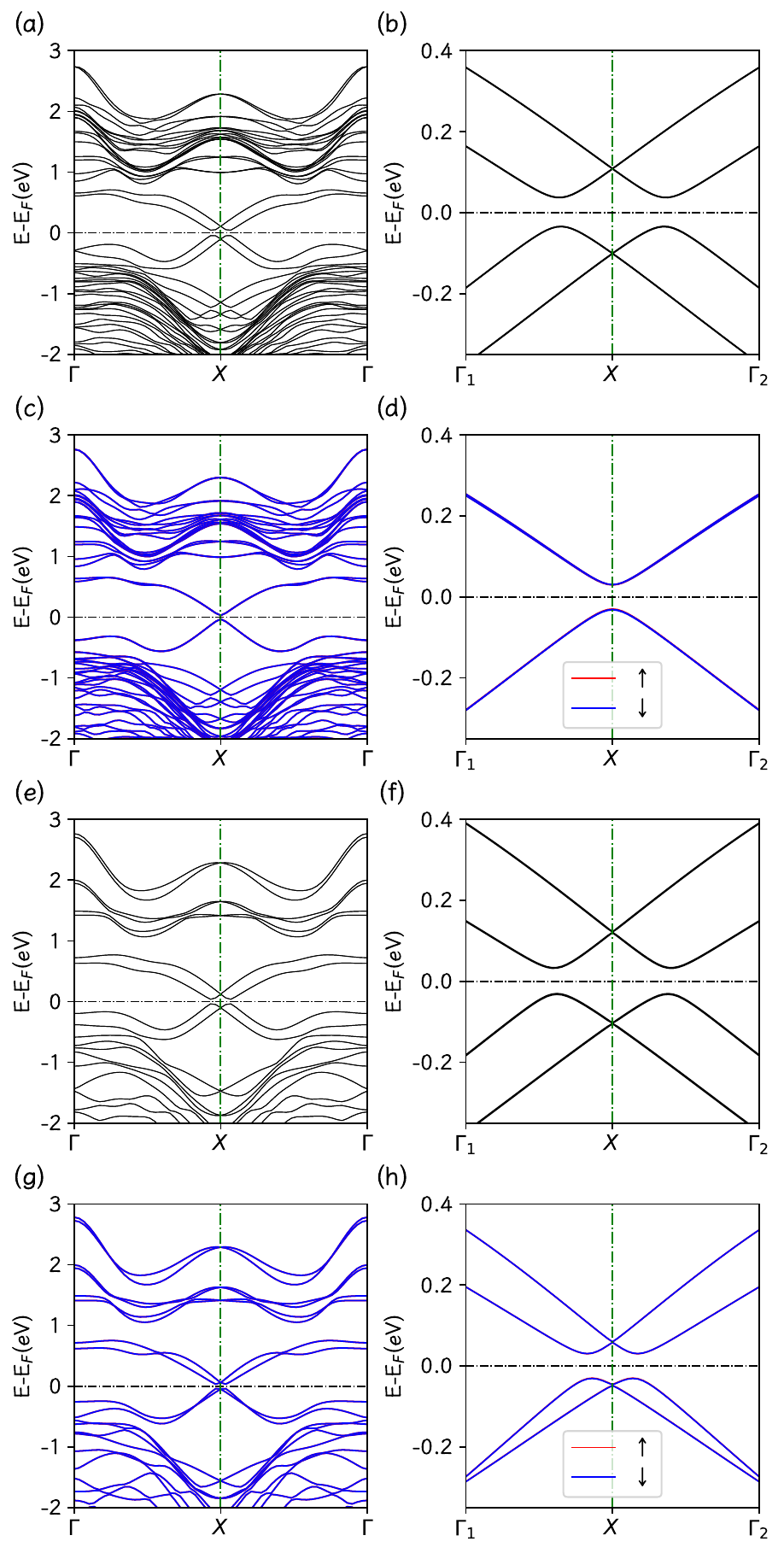}
\caption{\label{fig2} The band structures of the PtSe$_2$ nanoribbons with $N=10$ (the upper four panels) and $N=4$ (the lower four panels). The left-side panels describe the bands for the entire Brillouin zone, and the right-side panels those near the X point. The effect of SOC on the bands is demonstrated by showing the bands with (a, b, e, and f) and without (c, d, g, and h) taking SOC into account.}
\end{figure}

\subsection{Electronic band dispersion}

Our test calculation confirms that the monolayer PtSe$_2$ is a 2D semiconductor. We study the energy band structures of the PtSe$_2$ nanoribbons with $N=3\sim10$. The electronic structures of the PtSe$_2$ nanoribbons with $N=10$ and $N=4$, both with and without the SOC taken into account, are presented in Fig. \ref{fig2}. It is clear that there appears some new edge-related bands in the semiconductor gap of the monolayer PtSe$_2$. For the nanoribbon with $N=10$, the band structure without SOC is overall a  direct-gap semiconductor and the CBM and VBM are located at the X point in the Brillouin zone. It is very interesting that the conduction bands near the Fermi level can be well fit with a relativistic dispersion relation $E = \sqrt{\alpha_c p_k^2 + m^2}$ and the corresponding valence bands can be well fit with $E = -\sqrt{\alpha_v p_k^2 + m^2}$, where $p_k$ describes the k vector with respect to the X point and $m$ is half of the semiconductor gap. They are both double degenerate for edge and spin.

After taking SOC into account, there appears a large Rashba spin splitting near the X point in the edge-related bands, and both of the VBM and CBM are split into two astride the X point. This double unconventional Rashba splitting is different from those in a  parabolic band edge of conventional semiconductors\cite{rashba,rashba2,rashba3,sfet2003,psh2006prl,bitei,pst2017rmp,refadd1,zsh-bitei,refadd2,refadd3,refadd4}. The two branches of the bands are connected by the relation
\begin{equation}\label{sys1}
E_{i\downarrow}(k) = E_{i\uparrow}(k+Q_i),
\end{equation}
where $i=c$ ($i=v$) corresponds to the conduction (valence) bands, ($\uparrow$,$\downarrow$ describes the spin orientation, and $Q_i$ is the k difference between the two band extrema. This band symmetry is the same as $E_{i\uparrow}(k) = E_{i\downarrow}(-k)$, reflecting the time reversal symmetry. The  semiconductor gap is $E_g=0.072$ eV and the band dispersion relation read:
\begin{equation}\label{Ecv}
\left\{ \begin{array}{l}
\displaystyle E_{cs}=\sqrt{\alpha_c^2(p_k -\epsilon_s \gamma_c)^2 + m^2}\\
\displaystyle  E_{vs}=-\sqrt{\alpha_v^2(p_k -\epsilon_s \gamma_v)^2 + m^2},
\end{array}\right.
\end{equation}
where $s=\uparrow$ or $\downarrow$, and $\epsilon_s=+1$ or $-1$, respectively. Here, the parameters are obtained by fitting the calculated bands ($N=10$): $m=0.036$ eV, $\alpha_c=2.1$ eV\AA, $\alpha_v=2.2$ eV\AA, $\gamma_c=0.048$ \AA$^{-1}$, and $\gamma_v=0.043$ \AA$^{-1}$. The parameter $\gamma_c=Q_c/2$ ($\gamma_v=Q_v/2$) describes the Rashba k shift of the CBM (VBM), and the unconventional Rashba splitting energy is defined as $E_{Ri}=\sqrt{\alpha_i^2 \gamma_i^2 + m^2}-m$. They are comparable to those of BiTeI with giant Rashba spin splitting\cite{bitei,zsh-bitei}.

\begin{figure}
\includegraphics[width=0.82\columnwidth]{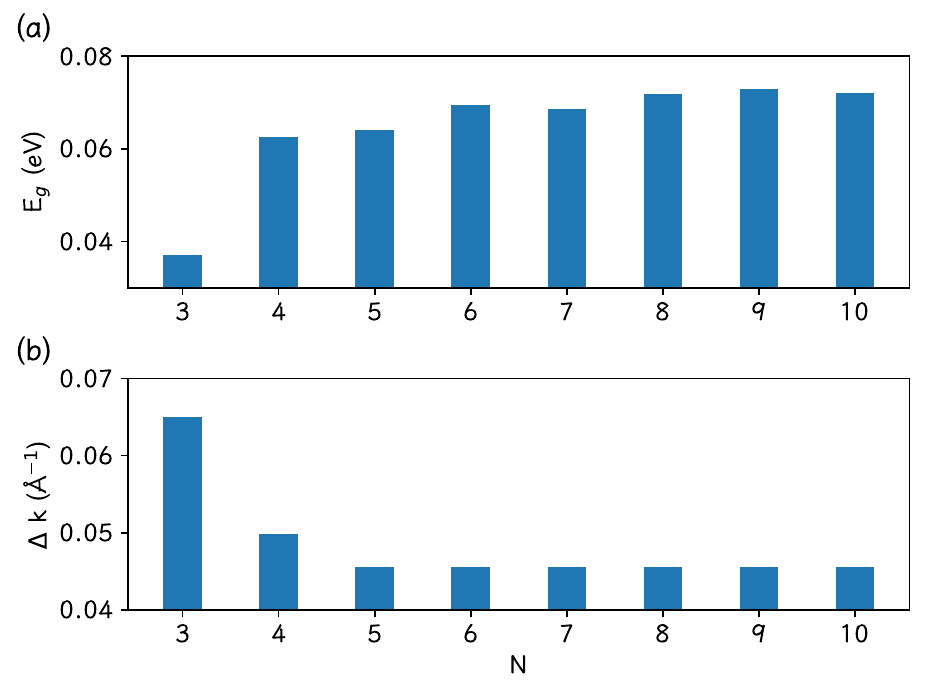}
\caption{\label{fig3} The $N$-dependent energy gaps ($E_g$ in eV, a) and average k splitting values ($\Delta k$, b) of the PtSe$_2$ nanoribbons, with SOC taken into account. $N$ describes the width of the nanoribbons.}
\end{figure}

\begin{figure*}
\includegraphics[width=0.82\textwidth]{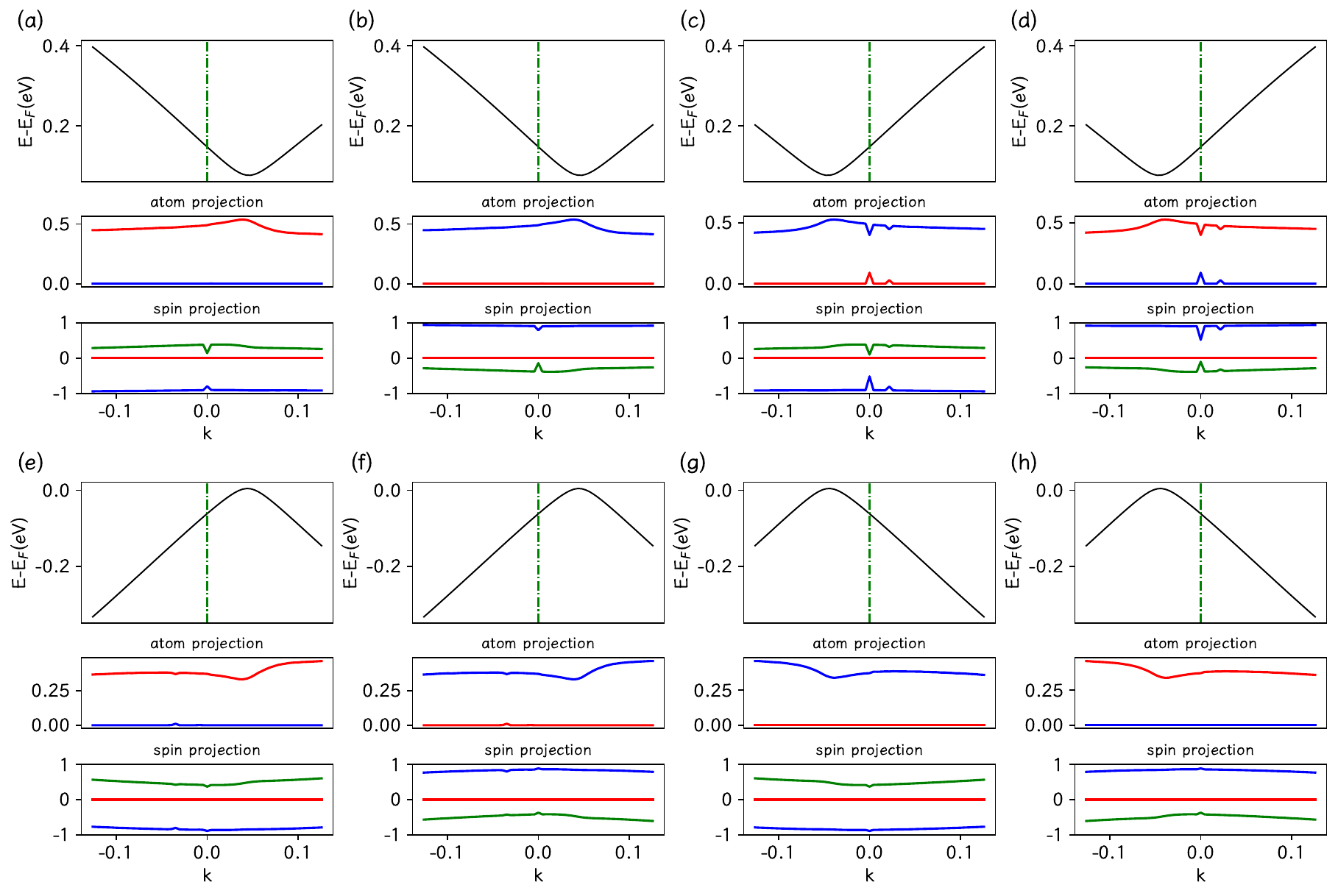}
\caption{\label{fig4} The edge and spin features of the eight edge-related bands near the X point for $N=10$ (a-h). For each of the bands, the atomic projection shows the weight of the Pt atom at the left edge (red line) or that at the right edge (blue line), and the spin projection  describes the variation of the three spin components along the band. For the spin projection, the red, green and blue means the $x$, $y$ and $z$ directions, respectively. The origin of the coordinate axis is set at the X point in the Brillouin zone.}
\end{figure*}

For $N=4$, the low-energy bands with SOC can be described well by Eq. (\ref{Ecv}), too.  Here, the parameters are $m=0.032$ eV, $\alpha_c=2.1$ eV\AA, $\alpha_v=2.2$ eV\AA, $\gamma_c=0.054$ \AA$^{-1}$, and $\gamma_v=0.045$ \AA$^{-1}$. It is clear that the low-energy bands with SOC changes little when $N$ switches from 10 to 4. In contrast, a big change happens in the low-energy bands without SOC. There appear large Rashba-like splitting in the energy bands near both CBM and VBM. This splitting for small $N=4$ can be attributed to the interaction between the two edges. It is interesting that this edge-based Rashba-like splitting is similar to the unconventional Rashba spin splitting due to SOC.

We present in Fig. \ref{fig3} the gap $E_g$ and $\Delta k=(\gamma_c+\gamma_v)/2$ as the main features of the $N$ dependent low-energy band structures. A trend can be seen that $E_g$ increases with $N$ and $\Delta k$ decreases with $N$. Most importantly, it appears that both $E_g$ and $\Delta k$ converge to constants and $\delta\gamma=|\gamma_c-\gamma_v|$ tends to zero when $N$ becomes large enough.

\subsection{Edge feature and Rashba spin splitting}

It is interesting to characterize the edge and spin feature in the low-energy bands.
First of all, it should be pointed out that in addition to the time reversal symmetry, there exists a space inverse symmetry for odd $N$, which makes the energy bands satisfy $E_{is}(k) = E_{is}(-k)$. In contrast, the bands for even $N$ are different, which is necessary to the unconventional Rashba spin splitting in the bands for even $N$. In the following, we shall concentrate our attention on the even $N$.

The local density of states (LDOS) is useful to characterize real-space distribution of atomic and orbital weight in the energy bands of the PtSe$_2$ nanoribbons. For $N=10$, we have investigated the LDOS of the low-energy bands (in the energy gap of the monolayer) of the nanoribbon. It is found that the LDOS is located mainly at the two edges and the LDOS reduces to zero in the middle of the nanoribbon. Therefore, the edge-related bands, especially those low-energy parts, can be attributed to the edges.

In order to make it clearer, we present in Fig. \ref{fig4} the key-atom and spin distribution of the low-energy parts of the eight bands near the X point in the Brillouin zone. It is clear that these atomic and spin weights do not change much along the bands near the X point, as shown in Fig. \ref{fig4}. Then, we can characterize the atomic weights by analysing the atom-resolved weights of the band edges (CBM and VBM). For $N=10$, the Pt and Se atoms in a computational unit cell of the nanoribbon can be numbered as Se1, Pt1, Se2, Se3, Pt2, Se4, $\cdots $, Se17, Pt9, Se18, Se19, Pt10, Se20. For one Se1+Pt1+Se2+Se3 unit, the summed weight of the conduction band at X is equivalent to 89.2\%, and that of the valence band reaches 96.3\%. The weight of one Se+Pt+Se unit actually jumps down and reaches zero for the fifth array of Se+Pt+Se unit from the edge. It is also clear in Fig. \ref{fig4} that the spin x component is zero, the y component is small, and the spin z component is near 1. It can be supposed that the true spin direction can be defined as the y-tilted z direction, and then the true spin z component is set to 1 and the true y component is also zero. Therefore, it is made clear for even $N$ that each of the eight low-energy Rashba-split bands is confined to one of the edges and connected with the true spin components (the z index is still used for the true z component.).

\subsection{SU(2) spin symmetry and persistent spin helices/textures}

With the characterization of the eight low-energy bands, we can divide the four conduction (valence) bands into two parts so that each of the parts is confined to one of the edges. Thus, we can write the following Hamiltonian for one of the edges.
\begin{equation}\label{ham1}
\begin{split}
\hat{H} = \sum_{k,s}\Bigl[\sqrt{\alpha_c^2(k+\epsilon_s\gamma_c)^2 + m^2} c_{kcs}^{\dagger}c_{kcs}\\
 - \sqrt{\alpha_v^2(k+\epsilon_s\gamma_v)^2 + m^2} c_{kvs}^{\dagger}c_{kvs}\Bigr]
\end{split}
\end{equation}
where the k vector is defined with respect to the X point in the Brillouin zone. For the other edge, the Hamiltonian has the same form with the opposite spin due to the time reversal symmetry.

It can be proved that $[\hat{H},c_{kis}^{\dagger}c_{kis}] = 0$ from this Hamiltonian (\ref{ham1}). We can construct the following SU(2) spin operators\cite{psh2006prl},
\begin{equation}\left\{
 \begin{array}{l}
 \displaystyle S^+_{iQ_i}=\sum_{k}c_{ki\uparrow}^{\dagger}c_{(k+Q_i)i\downarrow}, \\
 \displaystyle  S^-_{iQ_i}=(S^+_{iQ_i})^{\dagger},\\
 \displaystyle  S^z_{i}=\frac{1}{2}\sum_{k} (c^{\dagger}_{ki\uparrow} c_{ki\uparrow}-c^{\dagger}_{ki\downarrow} c_{ki\downarrow}),
\end{array}\right.
\end{equation}
where $i=c$ ($i=v$) corresponds to the conduction (valence) bands. It can be proved that these operators obey the following commutation relations.
\begin{equation}
[S^{z}_i,S^{\pm}_{iQ_i}]=\pm S^{\pm}_{iQ_i}, ~[S^+_{iQ_i},S^-_{iQ_i}]=2S^z_i
\end{equation}
\begin{equation}
\begin{split}
[\hat{H},S^{\pm}_{iQ_i}]=[\hat{H},S^z_{i}]=0
\end{split}
\end{equation}

\begin{figure}
\includegraphics[width=0.4\textwidth]{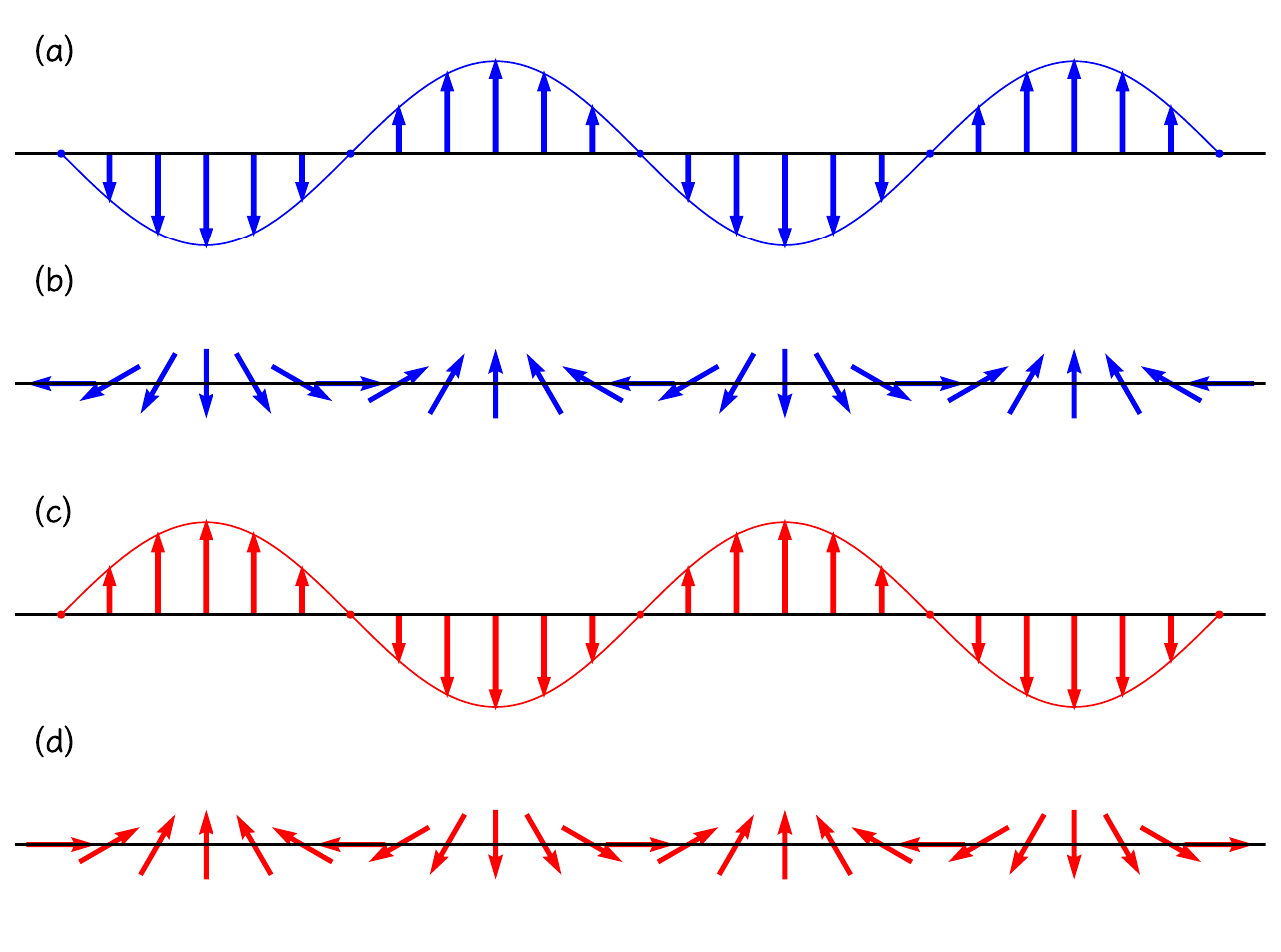}
\caption{\label{fig5} The persistent spin helixes for the conduction bands (a, b) and the valence bands (c, d). The x component is shown in (a) and (c), and the spin vector (in the x-y plane) is described in (b) and (d).}
\end{figure}

These results hold independently for CB and VB, and thus there are two series of the SU(2) symmetry for CB and VB, respectively. These commutation relation show the possibility of PSH in $[S^x(Q_c),S^y(Q_c)]$, which can lead to infinite lifetime of spin polarization\cite{psh2006prl,pst2017rmp}. As for the characteristic spatial periodicity of $l_{\text{PSH}} = \frac{2\pi}{|\bm{Q}|}$,  it can be shown that the $l_{\text{PSH}}$ is equivalent to 6.5 nm (7.1 nm) for the conduction (valence) bands in the case of  $N=10$, because of $Q_c=0.096$ \AA$^{-1}$ and $Q_v=0.088$ \AA$^{-1}$. In Fig. \ref{fig5}, we show these PSHs of the conduction and valence bands along one of the edges. This short pitch size of PSH can be useful in the future development of high-density scalable spintronic devices\cite{pst2017rmp,st2d2020rmp}.

Furthermore, we can demonstrate an effect of the  SU(2) spin symmetry on carriers in terms of quantum mechanics. Replacing the hole operator $h^{\dag}_{kvs}$ for the electron $c_{kvs}$, we can change Hamiltonian (\ref{ham1}) into the following form.
\begin{equation}\label{ham1h}
\begin{split}
\hat{H} =& \sum_{k,s}\Bigl[\sqrt{\alpha_c^2(k+\epsilon_s\gamma_c)^2 + m^2} c_{kcs}^{\dagger}c_{kcs}\\
& + \sqrt{\alpha_v^2(k+\epsilon_s\gamma_v)^2 + m^2} h_{kvs}^{\dagger}h_{kvs}\Bigr]+E_0
\end{split}
\end{equation}
It is clear that the ground state $| \rangle$ is defined by $c_{kcs}| \rangle=h_{kvs}| \rangle=0$. Suppose we have one doped electron and the lowest states can be described as $|k,s\rangle=c^{\dag}_{kcs}| \rangle$ with $k=\pm\gamma_c$. The electron can have the k vector $k=-\gamma_c$ and spin $s=\uparrow$, or $k=\gamma_c$ and $s=\downarrow$. For both of the cases, $|-\gamma_c,\uparrow\rangle$ and $|\gamma_c,\downarrow\rangle$, the spin z component is conserved and the electron moves with the constant speed.

If the electron takes the following superposition of these two states with a phase difference $2\lambda_j$,
\begin{equation}\label{fix}
|\lambda_j \rangle=e^{-i\lambda_j}|-\gamma_c,\uparrow\rangle+e^{i\lambda_j}|\gamma_c,\downarrow\rangle,
\end{equation}
then the spin z component is not conserved. It can be shown that the state $|\lambda_j \rangle$ with $2\lambda_j=0$ (or $\pi$) is the eigenstate of $S^x(Q_c)$ with the eigenvalue $1/2$ (or $-1/2$) (not for $S^y(Q_c)$). We can rotate $[S^x(Q_c),S^y(Q_c)]$ by $\theta_j$, and thus the $S^x(Q_c)$ is replaced by
\begin{equation}\label{Sx}
S^x(Q_c,\theta_j)=S^x(Q_c)\cos(\theta_j)+S^y(Q_c)\sin(\theta_j).
\end{equation}
Then, we can show that $|\lambda_j \rangle$ is the eigenstate of  $S^x(Q_c,2\lambda_j)$ with eigenvalue $1/2$ if we set $\theta_j=-2\lambda_j$. Letting $\lambda_j=\gamma_cR_j$, it is clear that  the x component $S^x(Q_c,2\lambda_j)$ in the rotated frame is conserved in terms of quantum mechanics. If one prepares the initial spin state in a given direction, one will observe a real-space $\cos$ pattern (persistent spin texture\cite{pst2017rmp}) along the direction and the PSH will be formed in the spin plane including the initial spin direction, as shown in Fig. \ref{fig5}. On the other hand, if we dope a hole in the valence bands, the same result can be obtained. These are consistent with experimental observation in high-mobility electron gases \cite{pst2017rmp,psh1qw,psh2qw,st2d2020rmp} and monolayer WTe$_2$\cite{psh3wte}.

\subsection{1D Dirac fermion system}

With the $N$ trend explained above, we believe that when $N$ is large enough, the symmetry between the conduction bands and the valence bands near the band edges will come true. Then the Hamiltonian (\ref{ham1}) can be written as
\begin{equation}\label{ham2}
\hat{H} = \sum_{k,j,s}\kappa_j\sqrt{\alpha^2(k+\epsilon_s\gamma)^2 + m^2} c_{kjs}^{\dagger}c_{kjs},
\end{equation}
where $\kappa_j=1$ ($\kappa_j=-1$) for $j=c$ ($j=v$), $\alpha=\alpha_c=\alpha_v$, and $\gamma=\gamma_c=\gamma_v$. The Hamiltonian (\ref{ham2}) can be considered to be the diagonalized version of the following one-dimensional Dirac Hamiltonian.
\begin{equation}\label{ham3}
\hat{H} = \sum_{k,s}\bar{c}_{ks}^{\dagger}\bigl[ \alpha(k+\epsilon_s\gamma)\sigma_x+m\sigma_z \bigr]\bar{c}_{ks},
\end{equation}
where $\bar{c}_{ks}^{\dagger}=(\bar{c}_{kcs}^{\dagger},\bar{c}_{kvs}^{\dagger})$, and $\sigma_x$ and $\sigma_z$ are the Pauli matrixes describing the conduction and valence bands near the band edges $(c,v)$. $\alpha$ is similar to the Fermi velocity $v_F$ of graphene\cite{rx00,rx01,rx02,st2d2020rmp}, but it is  approximately equivalent to $3.2\times 10^5$m/s ($\alpha\sim v_F/3$) and here we have a small finite mass $m/\alpha^2$. This Hamiltonian  (\ref{ham2} or \ref{ham3}) can describe the charge and spin of the pure Dirac fermion system in one dimension.

\section{CONCLUSION}

In summary, we have investigated the monolayer PtSe$_2$ nanoribbons with the stable neutral zigzag edges through combining first-principles investigation with low-energy model analysis. We find that the low-energy bands, in the semiconductor gap of the monolayer PtSe$_2$, can be well fit with relativistic energy dispersions in the energy window of [-0.2 eV, 0.2 eV] and there exists large unconventional Rashba spin splitting in the conduction and valence bands (for even $N$). Our analysis of atom-projected band structures and densities of states indicates that each of the low-energy bands originates from one of the nanoribbon edges. Our further investigation indicates that the semiconductor gap between the conduction band bottom and the valence band top will remain finite for large $N$. Most importantly, there exists SU(2) spin symmetry in both of the conduction and valence bands, which implies persistent spin helix (in the spin xy plane) and spin-conserving carrier transport (the z component) along each edge. Furthermore, when the inter-edge interaction becomes weak (with $N$ being large enough), a nearly-perfect Dirac fermion system  can be achieved through combining the valence and conduction bands. Thus we realize unconventional Rashba splitting with double SU(2) spin symmetry, persistent spin helices/textures, and  pure Dirac fermion systems in the stable monolayer PtSe$_2$ nanoribbons. They could be useful for future high-performance applications\cite{2Dtransistor1,2Dtransistor,futuretransistor,NiTe2,2Ddevice1,2Ddevice2}.\\


\begin{acknowledgments}
This work is supported by the Strategic Priority Research Program of the Chinese Academy of Sciences (Grant No. XDB33020100) and the Nature Science Foundation of China (Grant No.11974393) . All the numerical calculations were performed in the Milky Way \#2 Supercomputer system at the National Supercomputer Center of Guangzhou, Guangzhou, China.
\end{acknowledgments}



%

\end{document}